\documentclass{mn2e}
\usepackage[utf8x]{inputenc}
\usepackage{float}
\usepackage{graphicx}
\usepackage{amssymb}
\usepackage{amsmath}
\usepackage[usenames,dvipsnames]{xcolor}
\usepackage{astrojournals}
\usepackage{multicol}
\usepackage{multirow}
\usepackage[]{hyperref}
\usepackage{enumerate}
\usepackage{natbib}
\usepackage{placeins}
\usepackage{subfig}

\newcommand{\NoMod}{377 }

\title[Implications of the RSG and SN rate problems]{The red supergiant and supernova rate problems: implications for core-collapse supernova physics}
\author[S.~Horiuchi et al.]{S.~Horiuchi$^{1,2}$,\thanks{E-mail: horiuchi@vt.edu} 
K.~Nakamura$^{3}$,
T.~Takiwaki$^{4}$,
K.~Kotake$^{5}$,
M.~Tanaka$^{6}$\\
$^{1}$Center for Neutrino Physics, Department of Physics, Virginia Tech, Blacksburg, VA 24061, USA\\
$^{2}$Center for Cosmology, Department of Physics and Astronomy, University of California, Irvine, CA 92697, USA\\
$^{3}$Faculty of Science and Engineering, Waseda University, Ohkubo 3-4-1, Shinjuku, Tokyo 169-8555, Japan \\
$^{4}$Center for Computational Astrophysics, National Astronomical Observatory of Japan, Mitaka, Tokyo 181-8588, Japan \\
$^{5}$Department of Applied Physics, Fukuoka University, Fukuoka 814-0180, Japan \\
$^{6}$National Astronomical Observatory of Japan, Mitaka, Tokyo 181-8588, Japan \\
}
\begin{document}

\date{date}

\pagerange{\pageref{firstpage}--\pageref{lastpage}} \pubyear{2002}

\maketitle

%%%%%%%%%%%%%%%%%%%%%%%%%%%%%%%%%%%%%%%%%%%%%%%%
%%%%%%%%%%%%%%%%%%%%%%%%%%%%%%%%%%%%%%%%%%%%%%%%

\label{firstpage}

\begin{abstract}
Mapping supernovae to their progenitors is fundamental to understanding the collapse of massive stars. We investigate the red supergiant problem, which concerns why red supergiants with masses $\sim16$--$30 M_\odot$ have not been identified as progenitors of Type IIP supernovae, and the supernova rate problem, which concerns why the observed cosmic supernova rate is smaller than the observed cosmic star formation rate. We find key physics to solving these in the compactness parameter, which characterizes the density structure of the progenitor. If massive stars with compactness above $\xi_{2.5} \sim 0.2$ fail to produce canonical supernovae, (i) stars in the mass range $16$--$30 M_\odot$ populate an island of stars that have high $\xi_{2.5}$ and do not produce canonical supernovae, and (ii) the fraction of such stars is consistent with the missing fraction of supernovae relative to star formation. We support this scenario with a series of two- and three-dimensional radiation hydrodynamics core-collapse simulations. Using more than 300 progenitors covering initial masses $10.8$--$75 M_\odot$ and three initial metallicities, we show that high compactness is conducive to failed explosions. We then argue that a critical compactness of $\sim 0.2$ as the divide between successful and failed explosions is consistent with state-of-the-art three-dimensional core-collapse simulations. Our study implies that numerical simulations of core collapse need not produce robust explosions in a significant fraction of compact massive star initial conditions.
\end{abstract}


%%%%%%%%%%%%%%%%%%%%%%%%%%%%%%%%%%%%%%%%%%%%%%%%
\section{Introduction}

The association of core-collapse supernovae (SNe, including Types II and Ibc) with the death of massive stars is now firmly established both theoretically and observationally. Massive stars with initial mass between $10$ and $30 M_\odot$ evolve into red supergiants \citep[RSG; e.g.,][]{Levesque2006} with extended envelopes that give rise to their spectral identification as hydrogen-rich Type II SNe upon core collapse. Higher mass stars that shed their outer envelopes evolve into naked helium or Wolf-Rayet stars \citep[WR;][]{Crowther2007} that eventually explode as hydrogen-free Type Ibc SNe. The transition from RSG to WR is expected to be a gradual one in mass, since metallicity, rotation, and binary mass transfer also significantly affect evolution. Nevertheless, observational estimates of the initial mass required to evolve into a WR is $\sim 30 M_\odot$ \citep{Massey2000}, consistent with the maximum RSG mass. Theoretical studies also find the mass required for a single solar-metallicity star to shed its hydrogen envelope to be just above $25 M_\odot$ \citep{Heger2003,Eldridge2004}. The link between RSGs and Type IIP SNe has been firmly established by advances in archival imaging of pre-SN progenitors \citep{Smartt2009}. Although similar strategies have not yet positively identified the progenitors of Type Ibc SNe, studies indicate the need for a significant contribution of binary stars to the Type Ibc SN progenitor population \citep{Eldridge2013}. 

Advances have also revealed new puzzles. \cite{Smartt2009} identified what they termed the ``RSG problem'', which refers to the unknown fate of the most massive RSGs. Based on $20$ Type IIP SNe with pre-images, the authors statistically derived an upper mass for Type IIP SN progenitors of $16.5 ^{+1.5}_{-1.5} M_\odot$ ($\sim 21 M_\odot$ at $95$\%C.L.), which falls short of the mass range of RSG populations that extend up to $\sim 30 M_\odot$. Numerous explanations have been considered. For example, if the stellar initial mass function (IMF) is steeper, the null observation of massive RSG progenitors becomes statistically less significant \citep{Smartt2009}. Or, the loosely bound hydrogen envelopes of the most massive RSGs may become unstable and lost immediately prior to core collapse \citep[e.g.,][]{Yoon2010}. Pre-SN mass loss could also bias progenitor mass estimates to lower values due to insufficient dust correction (\citealt{Walmswell2012}, but see \citealt{Kochanek2012}). Or, stellar evolution may limit the RSG mass to $\sim 20 M_\odot$ \citep{Groh2013}. Finally, under certain conditions massive stars collapse to black holes (BH) with optically dark or faint ``failed SNe'' \citep[e.g.,][]{Woosley2012,Lovegrove2013}, which may occur in the upper RSG mass range \citep{Kochanek2008,Clausen2014,Kochanek2014a}. 

Another puzzle is the apparent dearth of observed cosmic SNe when compared to expectations from the cosmic star formation rate (SFR), termed the ``SN rate problem'' \citep{Horiuchi2011}. The mismatch is a factor of $\sim 2$ and consistently observed at all redshifts where cosmic SN rate data are available, except in the local $O$(10) Mpc regime \citep{Horiuchi2011,Botticella2012}. Possible explanations involve a large fraction of optically dim SNe \cite[][]{Horiuchi2011}, updates to the SFR calibrations \citep{Horiuchi2013}, and updates to dust corrections of SFR data \citep{Mathews2014}. 

The picture is therefore more complex than a simplistic one where all massive stars evolve as single isolated stars and explode as luminous SNe. In this Letter, we investigate the success or failure of massive-star core collapse by focusing on the RSG and SN rate problems within a common theoretical framework using the ``compactness'' parameter, which quantifies the mass density structure of the SN progenitor. The importance of the compactness is well documented \citep[e.g.,][]{Burrows1987,Fryer1999}, especially in predicting the success or failure of core collapse \citep{OConnor2011,Ugliano2012}. In particular, stars in the mass range $16$--$30 M_\odot$ have large compactness, thus providing a solution to the RSG problem \citep{Kochanek2014b}. In this Letter, we show that the failed SN scenario naturally provides an explanation of the RSG and SN rate problems, and we use a series of two-dimensional numerical hydrodynamics simulations of core collapse, as well as state-of-the-art three-dimensional simulations, to support this scenario.

The Letter is organized as follows. In Section \ref{sec:problems}, we discuss the RSG and SN rate problems within a common framework of compactness. In Section \ref{sec:sim}, we discuss our hydrodynamics simulations and implications for the RSG and SN rate problems. We conclude with discussions in Section \ref{sec:discussion}. 

%%%%%%%%%%%%%%%%%%%%%%%%%%%%%%%%%%%%%%%%%%%%%%%%
\section{Characterizing the problems }\label{sec:problems}
%%%%%%%%%%%%%%%%%%%%%%%%%%%%%%%%%%%%%%%%%%%%%%%%

The compactness encodes structural properties of the progenitor that critically affects whether the star will explode as a SN or not. It is defined as
\begin{equation} \label{eq:compactness}
\xi = \left. \frac{M/M_\odot}{R(M_{\rm bary}=M)/1000\,{\rm km}} \right\vert_{ t },
\end{equation}
where $R(M_{\rm bary}=M)$ is the radial coordinate that encloses a baryonic mass $M$ at epoch $t$. The compactness is a measure of the shape of the mass density profile surrounding the iron core: the slower the density profile drop with radius, the larger the compactness. A shallower density gradient results in a larger accretion rate and hence ram pressure that a SN shock must overcome in order to turn what is initially an implosion into an explosion. Progenitors with higher compactness are therefore more difficult to explode \citep[e.g.,][]{OConnor2011,Ugliano2012,Nakamura2014b}. 

We adopt $M_{\rm bary}=2.5 M_\odot$ at $t=0$, i.e., the pre-SN epoch. While other definitions are possible, we find that $2.5 M_\odot$ is sufficiently larger than the iron core mass but deep enough in the progenitor to affect the accretion rate during the critical moments of SN shock revival. As we demonstrate in Section \ref{sec:sim}, it shows a remarkable correlation with the qualitative result of the collapse, in particular in the vicinity of the parameter space for explosion failures. This is consistent with \cite{OConnor2011}, where the compactness defined by $2.5 M_\odot$ mass enclosed at time of bounce was used to distinguish between collapse to a neutron star (NS) or BH. \cite{Sukhbold2014} have shown how the compactness at the pre-SN epoch is just as useful as an indicator as the compactness derived at the time of core bounce. Therefore, we proceed to define the compactness at the pre-SN epoch and label this $\xi_{2.5}$.

%-----------------------------------------------------------------  
\begin{figure}
\includegraphics[width=80mm,bb = 0 40 520 530]{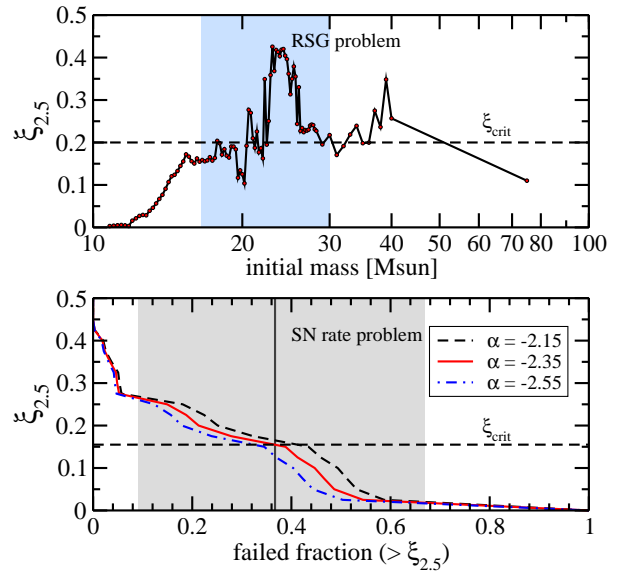}
\caption{{\it Top:} the RSG problem, showing the compactness as a function of initial stellar mass, for the solar metallicity pre-SN progenitors of WHW02. The shaded band indicates the mass range of the RSG problem, and the horizontal dashed line guides the eye to the necessary critical compactness needed to solve the RSG problem. {\it Bottom:} the SN rate problem, showing the compactness as a function of the failed fraction, defined as the fraction of massive stars with a compactness above a critical $\xi_{2.5}$ value. Three IMFs are shown as labeled. The shaded band indicates the missing fraction inferred from data, and the dashed line guides the eye to the required critical compactness.}
\label{fig:compactness}
\end{figure}
%----------------------------------------------------------------

The RSG problem in the language of compactness is illustrated in the top panel of Fig \ref{fig:compactness}, where $\xi_{2.5}$ is shown as a function of the initial progenitor mass, for the solar metallicity models of \cite{Woosley2002} (WHW02). The value of $\xi_{2.5}$ depends non-monotonically on mass, and in particular, a peak exists between $20$--$30 M_\odot$. The physics that determine the compactness is discussed in depth by \cite{Sukhbold2014}, and involves the detailed burning history of the star. The peak is driven partly by smaller mass progenitors forming degenerate compact white dwarf-like cores that drive the compactness lower, but also by the disappearance of the first carbon shell burning in progenitors above $\sim 20 M_\odot$ which increases the compactness since subsequent shell burnings are pulled down. While the exact features of $\xi_{2.5}$ are sensitive to the way semiconvection, convective overshooting, and mass loss are modelled, the qualitative peak features in mass are seen in multiple simulation codes \citep[][]{Sukhbold2014}. The shaded vertical column is the mass range of the RSG problem, i.e., $16.5$--$30M_\odot$ \citep{Smartt2009}: these stars are missing from Type IIP SN progenitor searches. Therefore, massive stars with compactness above a critical value $\xi_{\rm crit} \sim 0.2$ would need to not explode as canonically luminous Type IIP SNe. 

%-----------------------------------------------------------------  
\begin{figure}
\includegraphics[width=80mm,bb=-10 0 550 550]{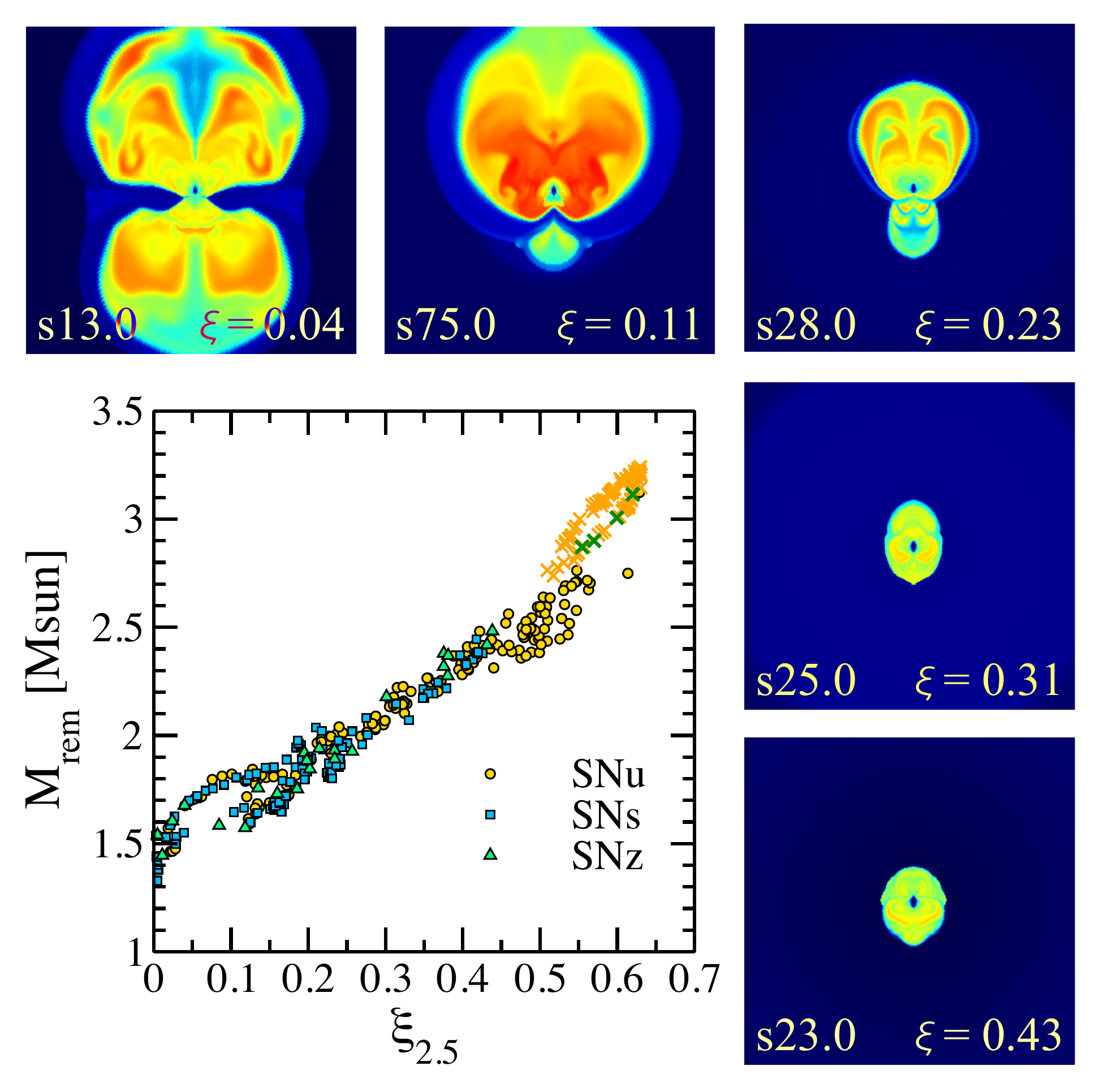}
\caption{Remnant mass versus compactness. Each point corresponds to a two-dimensional simulation with a different progenitor. A clear trend of failed explosions (crosses) populating high $\xi_{2.5}$ is observed. The five panels show the entropy at $200$ ms post bounce, from $5$ to $25 \, k_B/{\rm baryon}$ (blue--red). Each box is 2000 km on each side. Progenitors are chosen to cover a $\xi_{2.5}$ range $0.04$--$0.43$. }
\label{fig:correlation}
\end{figure}
%----------------------------------------------------------------

The SN rate problem can be framed in terms of $\xi_{2.5}$ by estimating the fraction of all massive stars that have $\xi_{2.5}$ above some critical value. Physically, this fraction is interpreted as massive stars whose core density profile make it difficult to explode as canonical SNe, and is equivalent to an optically dim or dark SN fraction. Defining massive stars as $8$--$100 M_\odot$ stars, we derive this fraction for three IMFs $dN/dM \propto M^{\alpha}$ with $\alpha = -2.15$, $ -2.35$, and $ -2.55$, shown in the bottom panel of Fig \ref{fig:compactness}. We quantify the SN rate problem using the latest data sets. The $z=0$ SFR density is $0.015 \pm 0.003  \, h_{70}  \, M_\odot {\rm  yr^{-1} Mpc^{-3}}$ and evolves as $(1+z)^{3.28}$ until $z\approx 1$ for the Salpeter-A IMF \citep{Hopkins2006,Hopkins2008}. Adopting a mass range of $8$--$100 M_\odot$ for stars producing SNe yields a SFR to SN rate conversion of $0.00965/M_\odot$ and hence a $z=0$ SN rate density of $(1.5 \pm 0.25) \times 10^{-2}  \, h_{70} {\rm  \,  yr^{-1} Mpc^{-3}}$. We use a set of the most recent 13 SN rate measurements  \citep{Dahlen2004,Cappellaro2005,Botticella2008,Bazin2009,Li2011,Dahlen2012,Melinder2012,Taylor2014}, and apply additional redshift-evolving dust correction as suggested in \cite{Mattila2012}, except to \cite{Dahlen2012} and \cite{Melinder2012} that already implement such corrections. Adding statistical and systematic uncertainties in quadrature, and fitting a function $\dot{N}_0 (1+z)^\beta$ to the data, we obtain $\dot{N}_0 = 0.92\pm0.15 \times 10^{-2}  \,  h_{70}^3{\rm  \, yr^{-1} Mpc^{-3}}$ and $\beta = 3.4^{+0.6}_{-0.4}$. The missing SN fraction and its uncertainty, both derived from data, are shown by the vertical band. A critical compactness of $ \sim 0.15$ therefore nominally solves the SN rate problem. Although the uncertainties are large, this is in the same range inferred to solve the RSG problem.

%%%%%%%%%%%%%%%%%%%%%%%%%%%%%%%%%%%%%%%%%%%%%%%%
\section{Simulations and implications } \label{sec:sim}
%%%%%%%%%%%%%%%%%%%%%%%%%%%%%%%%%%%%%%%%%%%%%%%%

We use numerical simulations of SNe to provide interpretations of the critical compactness identified in the previous section. Our two-dimensional models \citep{Nakamura2014b} are computed on a spherical polar grid of 384 radial zones from the center up to 5000 km 
and 128 angular zones covering $0 \leq \theta \leq \pi$, using the high-density EoS of \cite{Lattimer1991} with a nuclear incompressibility of $K = 220$ MeV. We employ the isotropic diffusion source approximation \citep[][]{Liebendorfer2009} and ray-by-ray approach to solve spectral transport of electron- and anti-electron neutrinos. Heavy-lepton neutrinos are treated by a leakage scheme to include cooling. We take into account explosive nucleosynthesis and its energy feedback into the hydrodynamics by solving a 13 $\alpha$-nuclei network \citep{Nakamura2014a}. Our three-dimensional models use a grid of 64 angular zones and 128 azimuthal zones (Takiwaki et al.,~in preparation).

The two-dimensional simulations are performed on the full set of solar- (s), $10^{-4}$ solar (u), and zero-metallicity (z) progenitors of WHW02, a total of $377$ progenitor initial conditions. This allows systematic trends to be explored. Fig \ref{fig:correlation} shows one such trend, showing the remnant mass, defined as the mass within a density coordinate of $10^{11} {\rm g cc^{-1}} $, as a function of the progenitor $\xi_{2.5}$, for all the s-, u- and z- models. Simulations that fail to reach a successful explosion, as defined by the shock radius not reaching a radius of 400 km within the first $1.5$ s of simulation time (equivalent to $\sim 1.2$ s in time since bounce), are marked by crosses. These failed explosions clearly populate the highest $\xi_{2.5}$ and confirm the predictive power of using $\xi_{2.5}$ to distinguish the outcome of core collapse. In fact, the compactness is a continuous measure of explosibility: the small panels of Fig \ref{fig:correlation} show how the explosion becomes increasingly smaller with higher $\xi_{2.5}$. 

Fig \ref{fig:correlation} implies that the division between successful (filled points) and failed (crosses) explosions is $\xi_{2.5} \sim 0.5$, i.e., a critical compactness that is significantly larger than the observationally inferred critical value of $\xi_{\rm crit} \sim 0.15$--$0.2$. However, there are issues that complicate such a conclusion. 
First, many of the successful explosions with large compactness will likely not explode as canonical SNe. For our EOS, the maximum NS mass is $\sim 2.4 M_\odot$ \citep[e.g.,][]{Muller2012}, so the remnants of progenitors with $\xi_{2.5} \gtrsim 0.4$ would collapse to BHs. Thus, these SNe will be dimmer due to mass fall back. Second, additional physics are considered to be important in accurately predicting the outcome of core collapse, including general relativity and weak interaction physics \citep[see recent reviews,][]{Kotake2012,Janka2012}. Also, studies suggest that stars explode more easily in two dimensions than in three (e.g., \citealt[][]{Hanke2012,Pejcha2012,Couch2013,Takiwaki2014}). All of these can shift the location of the critical compactness for a successful explosion. 

%-----------------------------------------------------------------  
\begin{figure}
\includegraphics[width=80mm]{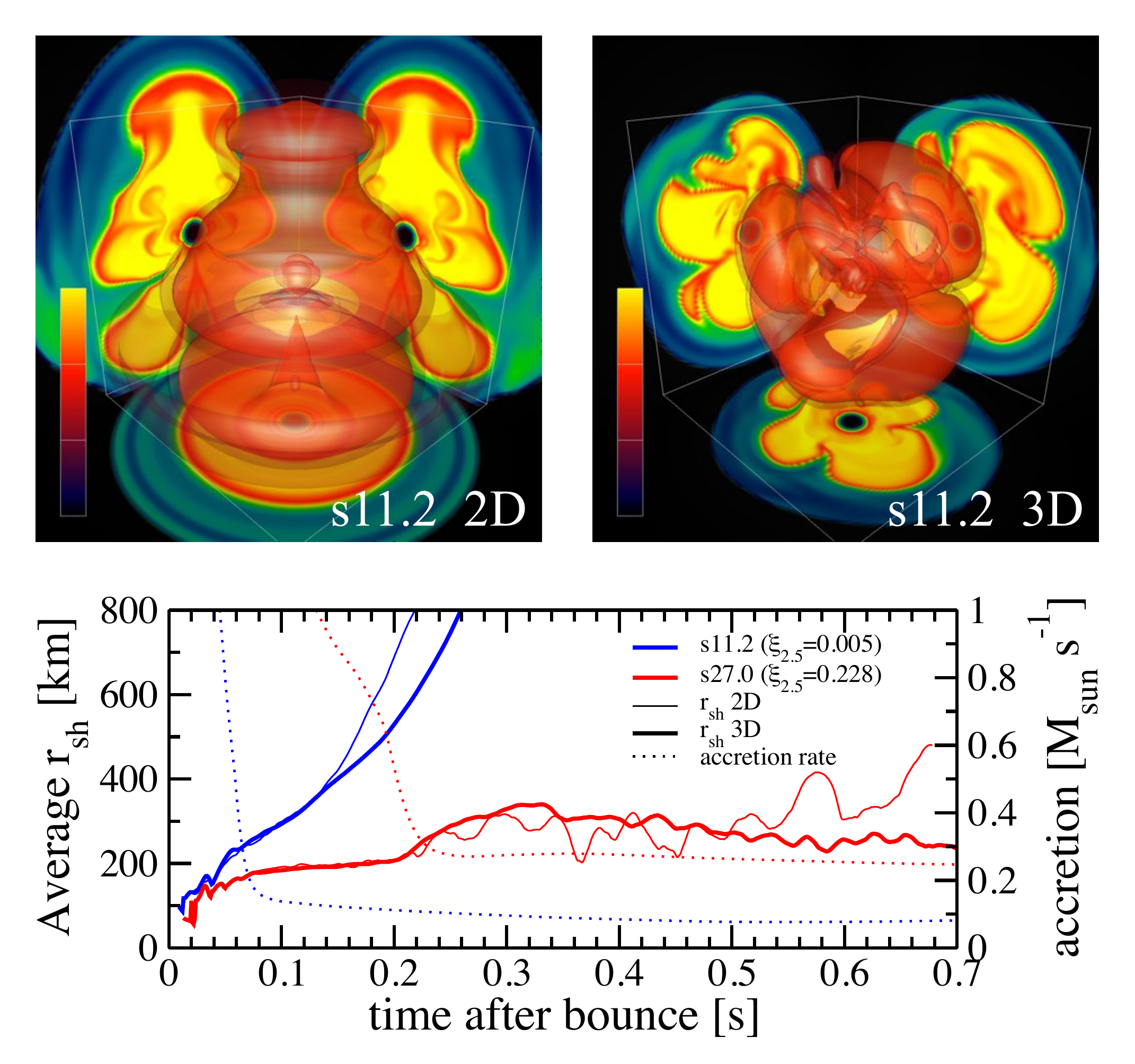}
\caption{ {\it Top}: snapshots of the entropy distribution of s$11.2$ at $200$ ms post bounce in 2D (left) and 3D (right), illustrating the smaller nature of 3D explosions. Each cube is 1000 km on each side, and the colour covers $5$--$16 \, k_B/{\rm baryon}$ (black--yellow). {\it Bottom}: average shock radius $r_{\rm sh}$ (solid line, left axis) for 2D (thin line) and 3D (thick line) simulations, and the mass accretion rate $\dot{M}$ (dotted line, right axis), all for two progenitors: s11.2 (blue) and s27.0 (red). 2D simulations experience earlier shock revival, and higher $\xi_{2.5}$ corresponds to later shock revival.}
\label{fig:threeD}
\end{figure}
%----------------------------------------------------------------

In order to provide a more realistic interpretation, we run select progenitors in both 2D and 3D. The lower panel of Fig \ref{fig:threeD} shows the average shock radius (left y-axis) for 2D (solid) and 3D (thick solid), and the mass accretion rate (dashed, right y-axis), for the s$11.2 M_\odot$ and s$27.0 M_\odot$ progenitor models of WHW02. The importance of $\xi_{2.5}$ is once again observable: while s$11.2 M_\odot$ ($\xi_{2.5} = 0.005$) comfortably explodes, s$27.0 M_\odot$ ($\xi_{2.5} = 0.228$) does not \citep{Takiwaki2014}. A more subtle but important point is that SN shock revival is slower in 3D than in 2D. This is clearly the case for s$11.2 M_\odot$, where the 3D shock radius grows slower than the 2D shock radius. For s$27.0 M_\odot$, the 3D shock radius shows no signs of revival, while the 2D shock does. These results are consistent with those of \cite{Hanke2013}. The upper panels of Fig \ref{fig:threeD} show the entropy distribution for the s$11.2 M_\odot$ progenitor in 2D (left) and 3D (right), both at $200$ ms post-bounce, illustrating how explosions are larger in 2D compared to the more round and compact 3D case. 

%-----------------------------------------------------------------  
\begin{figure}
\includegraphics[width=80mm,bb=0 40 520 300]{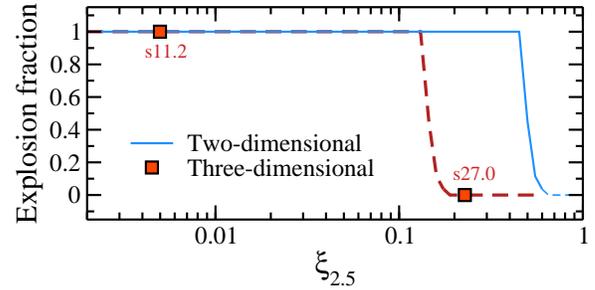}
\caption{The fraction of successful SN explosions as functions of the compactness $\xi_{2.5}$. Shown are the results from \NoMod 2D simulations (in solid) and results from two 3D simulations (square symbols). The dashed curve connects the 3D results whilst keeping the same shape as the 2D results. }
\label{fig:compare}
\end{figure}
%----------------------------------------------------------------

Fig \ref{fig:compare} illustrates the impact of dimensionality on the critical compactness. The solid curve shows the fraction of progenitors that successfully explode in our 2D simulations binned in $\xi_{2.5}$. The plateau and sharp decline in the explosion fraction illustrates the power of using $\xi_{2.5}$ as an indicator of the core collapse outcome. The square symbols show the results of 3D simulations. Each square corresponds to an individual progenitor as labeled. Interestingly, they support a quantitatively different picture, with the critical compactness being $\xi_{2.5} \sim 0.2$ or less. Only a limit can be stated due to the small number of progenitors simulated in 3D. Nevertheless, 3D simulations at present indicate it is possible for the critical compactness to be close to that required from observations. 

%%%%%%%%%%%%%%%%%%%%%%%%%%%%%%%%%%%%%%%%%%%%%%%%
\section{Discussions and conclusions}\label{sec:discussion}
%%%%%%%%%%%%%%%%%%%%%%%%%%%%%%%%%%%%%%%%%%%%%%%%

We have argued that current stellar and core-collapse simulations are consistent with the RSG and SN rate problems if all stars with compactness above a critical value of $\xi_{\rm crit} \sim 0.2$ collapse to form BHs and do not successfully produce canonically luminous SNe, i.e., failed SNe. We explicitly demonstrate how the compactness of the progenitor strongly impacts the outcomes of core collapse (Fig \ref{fig:correlation}), and that the required critical compactness (Fig \ref{fig:compactness}) is compatible with state-of-the-art simulations (Fig \ref{fig:compare}). 

There are several observational consequences of this scenario. In fact, similar values of $\xi_{\rm crit}$ have been previously inferred by consideration of remnant mass functions. \cite{Kochanek2014a,Kochanek2014b} show how the failed SN scenario reproduces the observed NS and BH mass functions without fine-tuning stellar evolution or core-collapse physics. 

Secondly, the implied failed SN fraction is $f_{BH}\sim20$--40\%, including stars in the $16.5$--$30 M_\odot$ range and additional contribution from stars around $\sim 40 M_\odot$. The value of $f_{BH}$ is presently only weakly constrained by data, but this is poised to improve. The diffuse SN neutrino background currently limits $f_{BH}\lesssim50$\% \citep{Lien2010}, but will improve with more data taking \citep{Horiuchi2009,Yuksel2012}. Campaigns searching for disappearance of massive stars will obtain $90$\% bounds down to $f_{BH} \sim 20$--30\% in 10 years observing \citep{Kochanek2008}. Galactic data also provide some probe \citep{Adams2013}. 

Thirdly, the scenario has implications for SN types and their progenitors. Consider a simple model where (i) $8$--$16.5 M_\odot$ stars produce IIP SNe, unless modified by binary interactions in which case they produce Ibc or IIb SNe, (ii) $16.5$--$30 M_\odot$ stars produce failed SNe, and (iii) $30$--$40 M_\odot$ stars produce Ibc or IIb SNe. We ignore the other Type II (IIn, IIL, etc) SNe here as they are a minority \citep{Li2011} and we assume them to be distributed across the full range of stellar masses \citep{Smith2012}. The IIP/(Ibc+IIb) ratio is then $0.015 (1-f_b)/(0.015 f_b + 0.0013)$, where $f_b$ is the binary fraction. The observed IIP/(Ibc+IIb) ratio $\approx 2.1$ \citep{Li2011} is then reproduced for a binary fraction $f_b \sim 0.3$, which is within current estimates based on O star statistics \citep[e.g.,][]{Sana2012}. This scenario predicts that $3/4$ of Ibc SNe arise from binary stripped stars of initial mass $<16.5 M_\odot$.

The main limitation of our study is the inherent uncertainties in the progenitor models that we have explored. The 1D progenitor models of WHW02 have the advantage of being the largest set of models currently available, but there are important uncertainties that must be highlighted. The treatment of turbulent transport and mixing inside the star is a source of significant uncertainty, and studies have begun to reveal just how important realistic treatments beyond the mixing-length style treatments are \citep[e.g.,][]{Meakin2011,Smith2014}. Notably, the 1D progenitor models of \cite{Limongi2006} do not show a peak in $\xi_{2.5}$ around $\sim 20 M_\odot$ seen in WHW02, although the number of progenitors is limited. On the other hand, \cite{Sukhbold2014} observe the peaks with the {\tt MESA} code. Further works are therefore required to reveal whether the peak is a robust feature of stellar evolution or not. 

More data and studies will also help reveal the severity of the RSG and SN rate problems and hence importance of our scenario. For example, dynamical modelling of SNe can help increase the number of SNe with progenitor mass estimates. At present, however, such mass estimates vary widely for any given SNe, highlighting the large systematic uncertainties involved. The significance of the SN rate problem will also be tested by future SFR data, and also as the systematics of SFR dust correction are studied further.

In summary, modern studies of SN progenitors collectively point to a non-negligible fraction of massive stars potentially failing to produce successful SNe. We have explored a connection of such observational features to the physics of core-collapse, namely, the compactness of the progenitors. We demonstrate how modern simulations support a trend of failed explosions at precisely where observations indicate potential failed explosions. An intriguing implication, then, is that the failures of many supernova simulations in producing robust explosions may actually be closer to reality than previously thought. 

%%%%%%%%%%%%%%%%%%%%%%%%%%%%%%%%%%%%%%%%%%%%%%%%
\vskip1cm
\noindent {\bf{Acknowledgments}} \\
%%%%%%%%%%%%%%%%%%%%%%%%%%%%%%%%%%%%%%%%%%%%%%%%

We thank John Beacom and Chris Kochanek for discussions and comments. SH is supported by a Research Fellowship for Research Abroad by JSPS. This study was supported in part by the Grants-in-Aid for the Scientific Research from the Ministry of Education, Science and Culture of Japan (Nos. 26870823, 24244036, 24103006, 26707013, 24740117, and 25103515) and by HPCI Strategic Program of Japanese MEXT, and used computational resources of the K computer at AICS through the HPCI project (ID:hp120304) and XC30 at NAOJ.

%%%%%%%%%%%%%%%%%%%%%%%%%%%%%%%%%%%%%%%%%%%%%%%%

\label{lastpage}

\end{document}